\documentclass[usenatbib]{emulateapj}
\usepackage{natbib}

\newcommand{\kms}{~\>{\rm km}\,{\rm s}^{-1}}
\newcommand{\mpch}{~\>h^{-1}{\rm {Mpc}}}
\newcommand{\degree}{^{\circ}}
\newcommand{\ud}{\,\rm{d}}

\newcommand{\lcdm}{\Lambda \rm {CDM}}
\newcommand{\om}{\Omega_{\rm {m}}}

\begin{document}

\title{Anomalous anisotropic cross-correlations between WMAP CMB
  maps and SDSS galaxy distribution and implications on the dark flow scenario}

\author{Zhigang Li\altaffilmark{1,2,*},
        Pengjie Zhang\altaffilmark{4,$\dagger$},
     Xuelei Chen\altaffilmark{1,3,$\ddagger$}  }

\altaffiltext{1}{Key Laboratory of Optical astronomy,
National Astronomical Observatories,Chinese Academy of
Science, Beijing 100012, China}

\altaffiltext{2}{Graduate University of Chinese Academy of Sciences, Beijing 100049, China}
\altaffiltext{3}{Center of High Energy Physics, Peking University, Beijing 100871, China}
\altaffiltext{4}{Key Laboratory for Research in Galaxies and Cosmology, 
Shang Astronomical Observatory, Nandan Road 80, Shanghai, 200030, China}

\altaffiltext{*}{zgli@bao.ac.cn}
\altaffiltext{$\dagger$}{pjzhang@shao.ac.cn}
\altaffiltext{$\ddagger$}{xuelei@cosmology.bao.ac.cn}

\begin{abstract}
We search for the dark flow induced diffuse kinetic Sunyaev Zel'dovich 
(kSZ) effect through CMB-galaxy cross correlation. Such angular 
correlation is anisotropic, with a unique $\cos(\theta_{\rm DF})$ angular 
dependence and hence can be distinguished from other components. Here, 
$\theta_{\rm DF}$ is the angle between the opposite dark flow direction 
and the direction of the sky where the correlation is measured. We analyze 
the KIAS-VAGC galaxy catalog of SDSS-DR7 and the WMAP seven-year 
temperature maps, applying an unbiased optimal weighting scheme to eliminate 
any statistically isotropic components and to enhance the dark flow 
detection signal. Non-zero weighted cross correlations are detected at 
$3.5\sigma$ for the redshift bin $z<0.1$ and at $3\sigma$ for the bin 
$0.1<z<0.2$, implying the existence of statistically anisotropic components in
CMB. However, further analysis does not support the dark flow explanation. 
The observed directional dependence deviates from 
the $\propto \cos\theta_{\rm DF}$ relation expected, 
and hence can not be explained by the presence of a single 
dark flow, and if the observed cross correlation is generated by the 
dark flow induced kSZ effect, the velocity would be too high ($\ga 6000$ km/s).
We report this work as the first attempt to search for dark 
flow through weighted CMB-galaxy cross correlation and to draw the 
attention on the sources of the detected anomalous CMB-galaxy 
cross correlation. 
 
\end{abstract}


\maketitle

\section{Introduction}      \label{introduction}

The peculiar velocity field of galaxies provides an important and robust 
way to understand 
the matter distribution of the universe. In the standard cosmology, the 
peculiar velocities arise from matter density perturbation and are unbiased 
tracer of the background matter distribution. It has been used to constrain the 
amplitude of matter power spectrum or the matter density fraction $\om$, and 
it could be used to detect possible sources of gravitational field which 
can not be observed directly with existing galaxy surveys. 

The peculiar velocity field of the nearby universe has been explored 
extensively using variant velocity tracers. 
 On small scales ($\leq20\mpch$) the peculiar velocities of galaxies are 
consistent with the standard $\lcdm$ model (\citealt{2005ApJ...635...11P,
2009MNRAS.400.1541A,2010JCAP...02..021J}). However, on larger scales,
various groups have reported bulk flows with unexpected amplitude. (1) On 
intermediate scales extending to about $100\mpch$, there is mounting evidence 
of a bulk flow with amplitude $\sim 400\kms$ using a composition of current 
velocity catalogs 
(\citealt{2009MNRAS.392..743W,2010MNRAS.407.2328F,2011MNRAS.tmp..391M}), 
which is unlikely within the frame of standard $\lcdm$ model, 
 despite the fact that there are some authors who claimed normal bulk 
velocities on these scales using seperate velocity catalog 
(\citealt{2011arXiv1101.1650N,2012MNRAS.420..447T,2012arXiv1208.2028M}). 
(2) On Gpc scales extending to $z\sim 0.3$, an even larger bulk flow of about 
$1000\kms$ was reported recently 
(\citealt{2008ApJ...686L..49K,2009ApJ...691.1479K,2010ApJ...712L..81K}), by
measuring the kinetic Sunyaev-Zel'dovich (kSZ) effect of galaxy clusters in
WMAP.  Whether or not this dark flow exists is still under intensive detate
(c.f. \cite{2011ApJ...737...98O} and \cite{2012arXiv1202.1339M} for
negative result, and \cite{2012arXiv1207.5338L} for peculiar velocities 
in $\lcdm$ simulation).  
If confirmed, this large bulk flow, dubbed ``dark flow'', would
severely challenge the standard model of cosmology and structure formation. 

It is therefore valuable to perform independent tests of the dark flow 
scenario beyond the cluster kSZ effect. In this paper we make an 
independent observational check of the dark flow scenario by 
cross-correlating the SDSS-DR7 galaxies with the WMAP 7 year CMB data.  
The galaxy-CMB cross correlation has been measured extensively 
(\citealt{2003ApJ...597L..89F,2006astro.ph.11046A,2006MNRAS.372L..23C,
2006PhRvD..74f3520G,2008PhRvD..77l3520G,2008PhRvD..78d3519H,
2008PhRvD..78d3520H,2010A&A...513A...3L,2008MNRAS.386.2161R,
2010MNRAS.402.2228S,2010MNRAS.406....2F,2010MNRAS.403.1261B,
2010A&A...520A.101H}), 
mainly to detect the integrated Sachs-Wolfe (ISW) effect. 
Similar analysis can be carried out to probe dark flow, under a specially 
designed weighing scheme.  As \cite{2010MNRAS.407L..36Z} (hereafter Z10) 
proposed, a large dark flow will  induce significant small scale anisotropies 
on the CMB temperature through the kSZ effect, due to fluctuations in 
the free electron number density. On each small patch of the sky, the 
kSZ effect induced by the clustering pattern of this dark flow traces 
the large scale structure closely. However, the overall amplitude is 
modulated and has a dipole pattern across the sky, with the positive 
maxima located at the direction opposite to the dark flow, and the 
negative minima located at the direction of the dark flow.  This 
statistically anisotropic distribution is distinctively different from 
all cosmic sources such as the primary CMB, the ISW effect, 
the thermal Sunyaev Zel'dovich (tSZ) effect, and the cosmic radio and 
infrared backgrounds. For this reason, one can apply a weighting function 
of appropriate directional dependence to eliminate the (statistically) 
isotropic components and isolate the kSZ component in CMB-galaxy cross 
correlations.  This is a major difference between our work and other
existing works on galaxy-CMB cross correlation.  
The proposed method can be extend to any
redshift to measure coherent large scale flow against CMB, if any. 
It is  complementary to the
recent measurement by \cite{2012arXiv1203.4219H}, which measures the mean
pairwise momentum of clusters through the kSZ effect of clusters and hence
probe the peculiar motion at smaller scales.   

In \S 2 we review the theoretical prediction of the dark flow induced
kSZ effect. The CMB temperature maps and galaxy catalog we used 
are described in \S 3, and the data analysis is done in \S 4.  
Results are shown in \S 5, along with investigations on possible 
systematics. We conclude the paper in \S 6, with a brief summary.

\section{The kSZ-galaxy cross correlation}
\label{theory}
Dark flow, if exists, will  induce CMB temperature fluctuations at all 
angular scales through the modulation of the inhomogeneous electron 
distribution on the uniform dark flow (Z10). The induced kSZ effect is 
given by\footnote{The original paper (Z10) missed a minus sign. However, 
numerical results presented in Z10 is unaffected. }
\begin{eqnarray}
\label{eq1}
\Theta^{kSZ}_T(\hat{r}) & \equiv & \frac{\Delta T^{kSZ}(\hat{r})}{T_{CMB}} = 
  \int n_{e}(\hat{r},z)\sigma_T \frac{-{\bf V}_{\rm DF}\cdot\hat{r}}{c} a \ud r \nonumber \\
          & = & \cos(\theta_{\rm DF})\times \int F_T(r)
  (1+\delta_e(\hat{r},z)) \ud r \nonumber \\
  &=& \cos(\theta_{\rm DF})\times \int F_T(r) \ud r \nonumber\\
  &+&\cos(\theta_{\rm DF})\times
    \int F_T(r) \delta_e(\hat{r},z) \ud r  \ .
\end{eqnarray}
where $T_{\rm CMB}=2.725K$ is the CMB temperature at present, 
${\bf V}_{\rm DF}$ is the dark flow velocity vector, with $V_{\rm DF}$ 
as its amplitude and $\theta_{\rm DF}$ the angle between the line of 
sight $\hat{r}$ and the {\it opposite} dark flow direction. The effective 
projection function for the kSZ signal is defined as 
\begin{equation}
F_T(r)\equiv \chi_e \sigma_T \Omega_b \frac{3H_0^2 }{8 \pi G} (1+z)^2
\frac{V_{\rm DF}}{c} \ ,
\end{equation}
where $\chi_e$ is the ionization fraction, $\sigma_T$ is the Thomson 
cross section, and $\Omega_b$ is baryon density parameter.
The integration in Eq.~(\ref{eq1}) is over the dark flow region along the 
line of sight. The first term in the last expression is a pure dipole term,
which has no correlation with the large scale structure. For this reason, we
will neglect this term hereafter. The second term represents the kSZ effect
arising from modulation of the inhomogeneous electron distribution
($\delta_e$) on the dark flow through the inverse Compton scattering. Clearly,
this component traces the large scale structure and is tightly correlated with
galaxy distribution. 
Hence it is promising to search for dark flow by cross correlating CMB
temperature maps with galaxy distribution over the redshift range of dark
flow. The induced cross correlation is 
\begin{equation}
w^{Tg}(\theta)\equiv \langle \Delta T^{\rm
  kSZ}(\hat{n})\delta_g^{\Sigma}(\hat{n}^{'})\rangle|_S\ .
\end{equation}
Here, $\delta^{\Sigma}_g$ is the galaxy surface number density fluctuation, 
$\theta$ is the angular separation between the two lines of sight, with 
$\cos{\theta}\equiv \hat{n}\cdot \hat{n}^{'}$. $w^{Tg}$ can be measured 
over either the whole survey area, or in each sub-region of it. We denoted 
this area average by a subscript ``S''. $\theta$ should not be confused 
with $\theta_{\rm DF}$, which is the angular separation between the 
{\it opposite} dark flow direction and the line of sight. 

The kSZ-galaxy cross correlations induced by dark flow have a number of 
signatures by which it can be distinguished from the cross correlations arising 
from other CMB components, such as the thermal Sunyaev Zel'dovich effect, 
the integrated Sachs-Wolfe effect, and the foreground contaminations. 
Below we discuss these one by one:

\begin{itemize}
\item $w^{Tg}$ is {\it statistically anisotropic}, with a characteristic
  $\cos(\theta_{\rm DF})$ angular
  dependence. On the contrary, cross correlations induced by the thermal SZ 
  effect, the ISW effect and extragalactic foregrounds are {\it statistically
  isotropic}. One can search for such signature by splitting the survey sky
  into sub-regions according to the
  values of $\cos(\theta_{\rm DF})$ and measure $w^{Tg}$ in each
  sub-regions. Alternatively, one can choose an appropriate weighting
  function $W(\hat{n})$ such that the weighted cross correlation
  \begin{equation}
w_W^{Tg}(\theta)\equiv \langle \Delta T^{\rm
  kSZ}(\hat{n})W(\hat{n})\delta_g^{\Sigma}(\hat{n}^{'})\rangle|_S
\end{equation} 
  vanishes for any isotropic CMB components. This requires $\langle W\rangle_S=0$. 
  Here we denote this weighted cross correlation
  with a subscript ``W''. The optimal weighting function minimizing the
  cross correlation measurement error is\footnote{Z10 proves that the optimal
  weighting function satisfies the equation $W=\langle
    W^2\rangle_S[\cos\theta_{\rm DF}-\langle \cos\theta_{\rm
        DF}\rangle_S]/\langle \cos\theta_{\rm DF}W\rangle_S$. Here we show
    that the solution to the above equation is $W=\cos\theta_{\rm DF}-\langle
    \cos\theta_{\rm DF}\rangle_S$, up to an arbitrary prefactor of no
    $\theta_{\rm DF}$ dependence.  }
\begin{equation}
\label{eqn:W}
W(\hat{n})=\cos\theta_{\rm DF}-\langle \cos\theta_{\rm DF}\rangle_S\ .
\end{equation}
The weighing scheme eliminates statistically isotropic components. 
Under this weighing scheme, the dark flow induced kSZ survives and 
the signal  is optimized. However, anisotropic components of more 
complicated angular dependence 
(e.g. $\cos(n\theta_{\rm DF})$ ($n=2,3,\cdots$)) also survive under the
weighting. So a non-vanishing $w_W^{Tg}$ means the existence of anisotropic
components, but not necessarily means the existence of dark flow. To further
test it, we will compare the measured $w^{Tg}$ as a function of
$\theta_{\rm DF}$ and check if it follows the $\cos(\theta_{\rm DF})$
dependence. 

\item $w^{Tg}$ should be frequency independent, since kSZ is frequency
  independent. 
\item $w^{Tg}_W(\theta)$ (and $w^{Tg}(\theta)$) should have similar $\theta$
  dependence as that of the galaxy angular correlation function $w^{gg}$. The
  two cross correlations are related by the following order of magnitude 
  approximation, 
\begin{eqnarray}
w^{Tg}_W(\theta)&\sim& f_W \tau_e \frac{V_{\rm DF}}{c} T_{\rm CMB}
\frac{b_e}{b_g}\times w^{gg}(\theta)\\
&\sim& 0.91\mu {\rm K} \times \frac{f_W}{1/3} \frac{\tau_e}{10^{-4}}\frac{V_{\rm
   DF}}{10^3 {\rm km}/s}\times w^{gg}(\theta)\nonumber\ .
\end{eqnarray}
Here, $f_W=\langle \cos(\theta_{\rm DF}) W(\hat{n})\rangle_S$, $\tau_e$ is the 
Thompson optical depth over the dark flow redshift range, $b_e$ is the bias of 
electron number distribution with respect to matter distribution and $b_g$ is 
the galaxy bias. For the dark flow direction ($l_{\rm DF}
=282^\circ$, $b_{\rm DF}=22^\circ$) reported by  \citet{2010ApJ...712L..81K},
$f_W \approx 0.09$ over the overlapping area of KIAS-VAGC and WMAP7. 
A complexity is that $b_e$ could have non-negligible scale dependence over
relevant scales due to various gastrophysics such as AGN and supernova
feedback. So the above relation only serve as a guideline. 
In the following, we use a fitting formula of gas window from 
hydrodynamical simulation of \cite{2006ApJ...640L.119J} to account for $b_e$.
We define the scaling factor as ratio of observed galaxy correlation function 
to the halofit model prediction (\citealt{2003MNRAS.341.1311S}), 
$R(\theta) = w^{gg}_{obs}/w^{gg}_{halofit}$, and the galaxy bias is simply,
$b_g = \sqrt{R}$. The resulted kSZ signal induced by a single dark flow are 
shown in Figure~\ref{ksz} as dashed lines.
\end{itemize}

Hereafter we will search for the dark flow through such signatures in the
WMAP-SDSS cross correlation. 

\section{Data}        \label{data}
Our cross correlation measurements are based on the KIAS-VAGC galaxy catalog 
of SDSS DR7 main galaxy survey \citep{2010JKAS...43..191C} and the CMB
temperature maps from the WMAP seven-year  data release. 
\subsection{KIAS-VAGC galaxy catalog}    \label{cagalog}
The main source of 
KIAS-VAGC is the New York University Value-Added galaxy catalog (NYU-VAGC) 
large scale structure (LSS) sample ``brvoid0'' \citep{2005AJ....129.2562B}, 
which has r-band magnitude limit of 17.6 rather than SDSS limit 17.7 to maintain 
a homogeneous survey depth and a simple survey boundary.
All magnitudes are k-corrected \citep{2003AJ....125.2348B} and evolved to rest 
frame magnitudes at redshift z=0.1 using an updated version of the evolving 
luminosity model of \cite{2003ApJ...592..819B}. The angular selection function 
is characterized carefully for each sector on the sky. The fiber collision 
results in about 7\% of targeted galaxies not having measured spectra. 
These galaxies are assigned the redshifts of their nearest neighbors with
available spectra. This method has been shown to be reasonably accurate for
large scale structure  
measurement \citep{2005ApJ...630....1Z}. Additionally, about 10k redshifts are 
borrowed from other spectroscopy galaxy surveys, such as the updated Zwicky 
Catalog (UZC) \citep{1999PASP..111..438F}, the IRAS Point Source Catalog 
Redshift Survey (PCSZ) \citep{2000MNRAS.317...55S}, the Third Reference 
Catalog of Bright Galaxies (RC3) \citep{1995yCat.7155....0D}, and the Two Degree 
field Galaxy Redshift Survey (2dF-GRS) \citep{2001MNRAS.328.1039C}. The resulted 
catalog contains 593,514 redshifts and covers 2.56 sr of the sky. 
In high density regions, KIAS-VAGC has high completeness. More details about 
the KIAS-VAGC catalog are given in \cite{2010ApJS..190..181C} and 
\cite{2010JKAS...43..191C}.

\begin{figure}
\resizebox{\hsize}{!}{\includegraphics{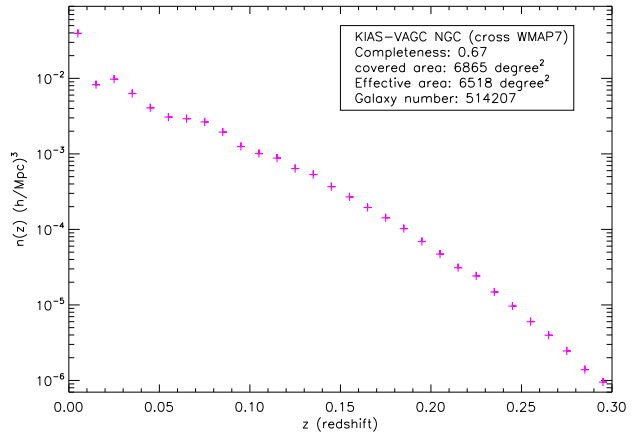}}
\caption{The galaxy number density with redshift in our home catalog.}
\label{nz}
\end{figure}

In this work we use the major part of KIAS-VAGC, the north Galactic cap (NGC), 
which covers an effective area of 6518 square degree. The sky area overlapping
with the WMAP7  temperature maps out of the KQ75 mask contains 514207
galaxies. The number density of the resulted galaxy catalog is shown in Figure 
\ref{nz}. To study the radial depth over which the dark flow extends, we split 
the galaxies into two redshift bins: $z<0.1$ and $0.1<z<0.2$.

\begin{figure}
\resizebox{\hsize}{!}{\includegraphics{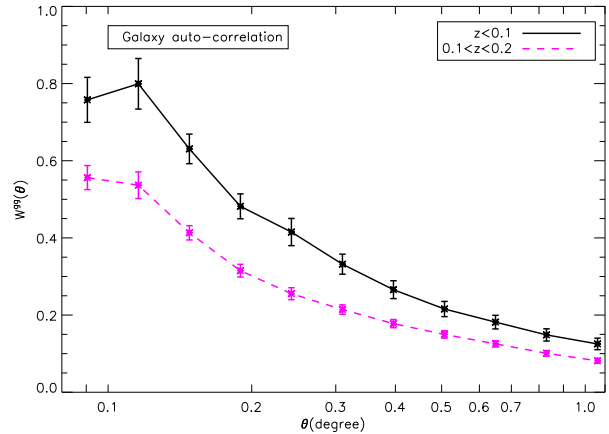}}
\caption{Auto-correlation functions of galaxies for $z<0.1$ (solid line) 
and $0.1<z<0.2$ (dashed line).} 
\label{galaxyauto}
\end{figure}

\subsection{CMB Temperature Maps}    \label{cmbmap}
The WMAP seven-year data products are publicly available in the Healpix 
format (\citealt{2005ApJ...622..759G}). The WMAP observed in 5 frequency 
bands, i.e. the $K$ band at 23~GHz, $Ka$ band at 31~GHz, $Q$ band at 41 
GHz, $V$ band at 61 GHz, and $W$ band at 94 GHz. The first two low 
frequency bands are used to extract Galactic foregrounds.
The 3 high frequency bands $Q$, $V$ and $W$ (with 8 channels 
totally) has subdominant foreground contaminations and relatively higher
resolutions, suitable for testing frequency-dependent contaminations to 
the small scale kSZ-galaxy cross correlations. 
In fact we do not see any considerable difference in the cross correlation 
results and the associated measurement errors of these 
three bands. We do not use the foreground-free ILC map provided by the 
WMAP team \citep{2009A&A...493..835D}, due to its low angular resolution 
(with an FWHM about $1\degree$). We use the CMB temperature maps with 
resolution 9 ($res=9$) in Healpix format, corresponding to an angular 
resolution of 7 arcmins. This results in 3145728 pixels with 47.21 square 
arcmin each. We use a combination of the ``Extended temperature analysis 
mask'' (KQ75) and the ``Point Source Catalog Mask'' 
(\citealt{2011ApJS..192...15G}) to exclude unobserved or foreground 
contaminated regions. About $70\%$ sky regions are available after these 
maskings.

To explore the systematics coming from intrinsic CMB fluctuations, and 
to test the robustness of the Jackknife error estimation method in this 
analysis, we generated 1000 CMB temperature simulations with the 
``Synfast'' code provided by WMAP. These simulated maps are drawn with 
the best-fit CMB power spectrum, then we add thermal noise to each map. 
The thermal noise are Gaussian distributed with a variance inversely 
proportional to the observation times of each pixel. The thermal noise 
in different pixels are uncorrelated.

Due to large errors in the cross correlation measurements, we do not use 
the cross-correlated mock samples for galaxy and CMB. It has been shown 
that the estimated errors in the cross correlation measurements using 
the CMB simulation maps are consistent with those using cross correlated 
mock samples with high precision (\citealt{2007MNRAS.381.1347C}).

\section{Cross correlation and error estimation}     \label{method}

The galaxy overdensity in the $i$-th pixel is 
$\delta_{g,i}^{\Sigma} = n_{g,i}^{\Sigma}/\bar{n}_{g,i}^{\Sigma} - 1$, 
where $n_{g,i}^{\Sigma} $ is the galaxy surface number density 
and $\bar{n}_{g,i}^{\Sigma}$ is the expected mean galaxy number density. 
The CMB temperature anisotropy is $\Delta T_{,i} = T_i - T_{\rm CMB}$. 
The cross correlation function at angular separation $\theta$ is given by

\begin{equation}
w^{Tg}(\theta) =\frac{ \sum_{i,j} [f_{g,i} \delta_{g,i}^{\Sigma}][ f_{T,j} 
              \Delta T_{,j}] }{ \sum_{i,j} f_{g,i} f_{T,j}} \ .   \label{eqwobs}
\end {equation}
Here $f_{g,i}$ is the galaxy selection function or completeness and $f_{T,i}$ 
is the fraction of pixel $i$ outside of the CMB masks. They are used to 
correct for the incompleteness of the survey. The sum is over all pairs 
whose separations are within the angular bins. 
Correspondingly, the weighted cross correlation is given by
\begin{equation}
w_W^{Tg}(\theta) =\frac{ \sum_{i,j} [f_{g,i} \delta_{g,i}^{\Sigma}][ f_{T,j}W_j 
              \Delta T_{,j}] }{ \sum_{i,j} f_{g,i} W_jf_{T,j}} \ ,
\end {equation}
where, $W_j\equiv W(\theta_j)$ is the weighting function given by Eq.(\ref{eqn:W}). 

Since we are interested in angular separation of degree scale
($\theta=O(1^{\degree})$), we can perform the above measurement for each small
patch of the sky, as long as the sky patch is much larger than one square
degree. $\theta_{\rm DF}$ is defined as the angle between the center of such
sky patch and the opposite dark flow direction. For statistically isotropic CMB
sources (tSZ, the ordinary kSZ, ISW, etc.) and cosmic radio and infrared
foregrounds, the induced cross correlations (contaminations for our purpose)
should be isotropic and be independent of $\cos\theta_{\rm DF}$. Only the dark
flow induced kSZ has the characteristic $\cos\theta_{\rm DF}$ angular
dependence. This is the key to separate this  component from other
correlations. 


We estimate the errors of the cross correlation measurements using the Jackknife 
method. We divide the joint survey region into 1000 subregions which have equal 
effective area (weighted by galaxy completeness). Then the Jackknife samples are 
built by discarding one of the subregions each time. The cross correlation is 
estimated by $\hat{w} = w_0 - (1-1/N) \sum_{i=1}^N w_i$, where $N$ is the number 
of subregions, $w_0$ is the cross correlation measured from the full survey 
region and $w_i$ from the $i$th Jackknife sample. This estimator is nearly bias 
free in general cases. Especially if $w$ is only a quadratic function  as this
case, the $\hat{w}$ defined here will be unbiased. The error of this estimator 
is given by 
\begin{equation}
\sigma^2(\hat{w}) = \frac{N-1}{N} \sum_{i=1}^N (w_i - \bar{w})^2  \ , 
\label{eqerr}
\end{equation}
where $\bar{w} = \sum_{i=1}^N w_i / N$ is the average of $w_i$.
To explore the cross talk between different angular bins, we also construct the 
full covariance matrix for the Jackknife samples,

\begin{equation}
C_{ij} = \frac{N-1}{N} \sum_{m=1}^N (w_m(\theta_i) - \bar{w}(\theta_i)) 
          (w_m(\theta_j) - \bar{w}(\theta_j)) \ . 
\end{equation}

\section{Measurements and implications}   \label{result}
\begin{figure}
\resizebox{\hsize}{!}{\includegraphics{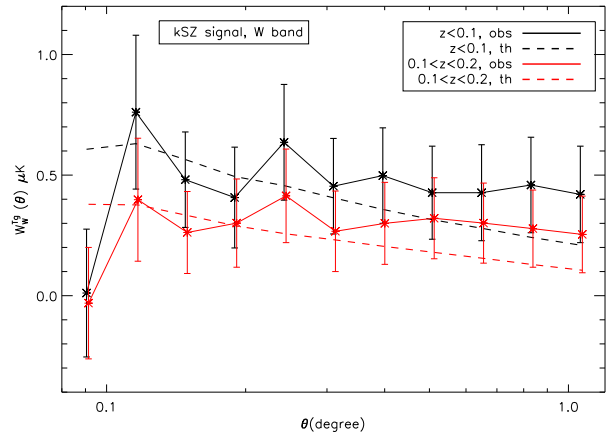}}
\caption{The weighted cross correlation of SDSS-DR7 galaxies and WMAP7yr CMB. 
The direction of the dark flow in the weighting is assumed to be $l=282^\circ$, 
$b=22^\circ$. The black solid line shows result for $z<0.1$, and red solid 
$0.1<z<0.2$. The black dashed line shows the kSZ signal predicted by a single 
dark flow of $9530\kms$ in redshift range $z<0.1$, and red dashed line 
a dark flow of $6140\kms$ in redshift range $0.1<z<0.2$.}
\label{ksz}
\end{figure}

\begin{figure}
\resizebox{\hsize}{!}{\includegraphics{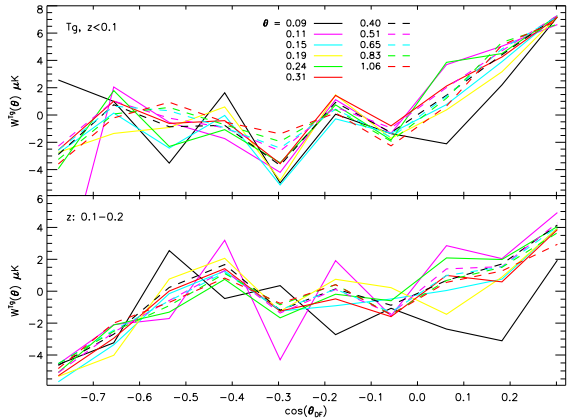}}
\caption{The $\cos \theta_{\mathrm{DF}}$ dependence of unweighted cross 
correlations, $w^{Tg}(\theta|\theta_{\mathrm{DF}})$, for redshift range 
$z<0.1$ (upper panel) and $0.1<z<0.2$ (lower pannel).
The $w^{Tg}(\theta|\theta_{\mathrm{DF}})$ in the angular bins ($\theta$) 
are shown as colored lines.
As shown, $w^{Tg}(\theta|\theta_{\mathrm{DF}})$ shows a complicated 
dependence on $\cos \theta_{\rm DF}$. However, on sub-regions with 
$\cos \theta_{\rm DF} > -0.3$ for low redshift bin and 
$\cos \theta_{\rm DF} < -0.4$ for high redshift bin, 
$w^{Tg}(\theta|\theta_{\mathrm{DF}})$ are nearly linear function of 
$\cos \theta_{\rm DF}$. It is also the case for high redshif sample 
on sub-regions with $\cos \theta_{\rm DF} > -0.3$, except that the 
linear relation is weak and noisy.
}
\label{gtsub}
\end{figure}


The weighted CMB-galaxy cross correlations are shown in figure~\ref{ksz}, 
for two redshift bins $z<0.1$ and $0.1<z<0.2$. The measurement is pretty
noisy. Nevertheless, we detect non-zero correlations at $3.5\sigma$ for 
low redshift bin and $3\sigma$ for high redshift bin. 

To further quantify the anisotropy of the CMB-galaxy cross correlation, 
we divide the survey region into 10 subregions according to the value of 
$\cos \theta_{\rm DF}$ and then measure $w^{Tg}$ in each sub-regions. The
results are shown in figure~\ref{gtsub}. The measured 
$w^{Tg}(\theta|\theta_{\mathrm{DF}})$ shows significant dependence on 
$\theta_{\mathrm{DF}}$, which is consistent with a non-zero $w^{Tg}_W$ 
signal. 

As explained earlier, the ensemble average of the weighted cross 
correction vanishes for statistically isotropic CMB components and 
galaxy distributions. Hence the detected non-zero cross correlation 
may imply either (1) statistical fluke caused by statistical fluctuations 
in CMB and galaxy distribution, (2) anisotropic and correlated
residual contaminations in CMB and galaxy distribution, or (3) the dark 
flow induced kSZ effect that we are searching for.

\subsection{Statistical fluke?}
We have performed a number of tests to check whether the detected 
abnormal anisotropic correlation is caused by statistical fluctuations 
in the intrinsically isotropic CMB and galaxy distribution. In particular, 
we pay attention to the intrinsic CMB temperature fluctuations, the galaxy 
shot noise and the CMB monopole and dipole. These components are uncorrelated 
with the galaxy distribution. However, they do induce statistical fluctuations 
to the measured CMB-galaxy cross correlations and can in principle mislead 
the interpretation of the measurements. 

\subsubsection{CMB fluctuations: intrinsic and secondaries}
Firstly we have generated 1000 simulated CMB temperature maps using the 
best fit angular power spectrum and the beam function provided by the 
WMAP team. Then Gaussian distributed white noise are added to each pixel 
of the map according to the observation times of that pixel. We have not 
included the 1/f instrumental noise since it has been removed during 
map-making procedure and it does not correlate with galaxies and 
hence only contributes statistical errors. The weighted cross correlation 
are measured for these simulated maps, and the results are shown in the 
upper panel of figure~\ref{syst}. As expected, the intrinsic CMB 
anisotropies do not bias our kSZ signal and the errors are consistent 
with those of Jackknife method.

The CMB secondaries, such as thermal SZ and ISW effect, in principle, 
are statistically isotropic. So under the designed weighting scheme, 
they contribute only the statistical error to our cross correlation 
measurements, rather than bias them. Simple order of magnitude 
estimation shows that the induced statistical fluctuations are 
negligible to that of primary CMB. The induced statistical error per 
multiple $\ell$ bin is $\propto \sqrt{(C^{\rm CMB}_\ell+C^{\rm tSZ}
_\ell+C^{\rm ISW}+C^{\rm inst}_\ell+\cdots)_W(C^g_\ell+C^{g,shot}
_\ell)/(2\ell+1)}$. Here, $C_\ell$ is the corresponding power spectrum 
of the primary CMB, the thermal SZ effect, the ISW effect, instrumental 
noise, galaxy number density and galaxy shot noise, respectively. The 
subscript ``$W$'' denotes the corresponding quantity under the anisotropic 
weighting scheme. At angular scales accessible to WMAP, primary CMB is 
the dominant component and hence contributes most to the statistical 
error. So for estimating the statistical error of the cross correlation, 
we can neglect the other components. Nevertheless, to fully consider 
these effects one needs to simulate the CMB temperature and galaxies 
simultaneously with cross correlations predicted by the corresponding 
components. While \cite{2007MNRAS.381.1347C} have concluded that within 
the 10\% level, the statistical errors coming from simulated CMB maps 
with and without cross correlations with galaxies are fully consistent, 
especially for scales smaller than $10\degree$. This result confirms 
our estimation. So in this work we just ignore the CMB secondaries in 
the statistical error estimation.

\subsubsection{Galaxy shot noise}
\begin{figure}  
\resizebox{\hsize}{!}{\includegraphics{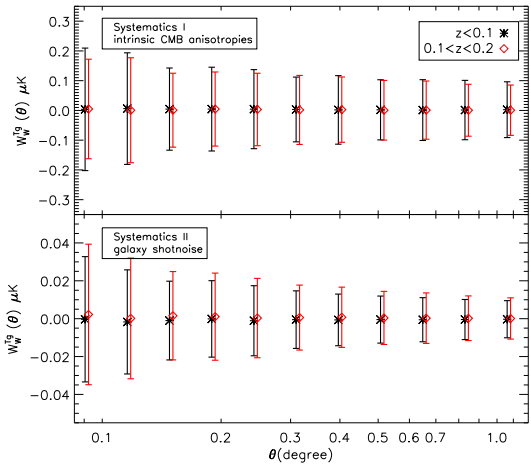}}
\caption{The systematics coming from intrinsic CMB anisotropies (upper) 
and galaxy shot noise (lower).} 
\label{syst}
\end{figure}

Secondly, we examine the galaxy shot noise effect by generating 100 
mock galaxy catalogs using the KIAS-VAGC selection function. Each mock 
catalog has the same number of galaxies as the real one. We have 
measured the weighted cross correlation for the mock catalogs and the 
real CMB data. The results are shown in the lower panel of 
Figure~\ref{syst}. Again, we do not see any bias effect and the errors 
are small enough to be neglected. We do not attempt to include the 
intrinsic galaxy clustering effect, since \cite{2007MNRAS.381.1347C} 
have shown that using simulated CMB maps one can get sufficient precision 
for this kind of cross correlation measurements with low signal-to-noise 
ratio.

\subsubsection{CMB monopole and dipole}
\begin{figure} 
\resizebox{\hsize}{!}{\includegraphics{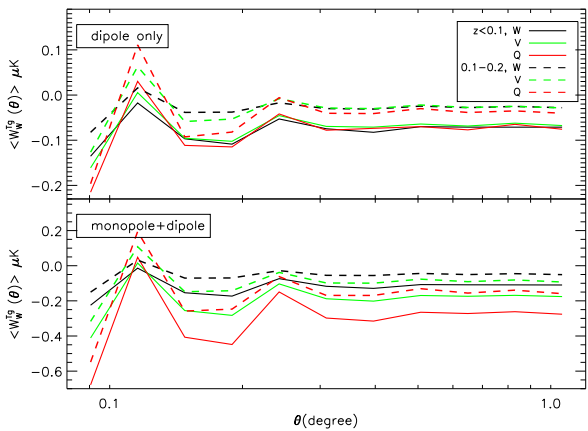}}
\caption{The effect of CMB dipoles and monopoles. The CMB dipoles and 
monopoles are extracted from the real maps using the Healpix subroutine 
``remove\_dipole''. The top panel shows the weighted cross correlations 
of galaxies and CMB dipoles, and bottom CMB dipoles+monopoles. }
\label{dipole_ksz}
\end{figure}

The CMB dipole are anisotropic over large angular scales, which could
interfere with our searching for dark flow. The CMB monopole has a large
amplitude, which may induce large statistical fluctuations in the cross
correlation. Hence we remove them  in the CMB maps we used. Nevertheless,  
there could exist residual  monopole and dipole, which may contaminate 
our cross correlation  measurement. To test this possibility, we cross 
correlate the monopole/dipole components of the CMB maps we used with 
the galaxies. The results for the three WMAP bands are shown in 
Figure~\ref{dipole_ksz}. The frequency dependence presented in the 
figure comes from the difference of dipoles on the maps. For the W 
band we are focused on, the residual dipole contribution to the weighted 
cross correlation is less than $0.1 \mu{\rm K}$, so is the residual 
monopole.

\subsection{Systematical errors from residual contaminations}
Based on these tests, we conclude that the measured anisotropic 
correlation is unlikely caused by statistical fluctuations, such as CMB 
primaries or the even smaller components, TSZ and ISW effects. 
Rather, it implies either a dark flow induced kSZ effect, or 
caused by the residual contaminations in WMAP/SDSS.

\subsubsection{Are CMB intrinsic fluctuations and galaxy distribution isotropic?} 
\begin{figure} 
\resizebox{\hsize}{!}{\includegraphics{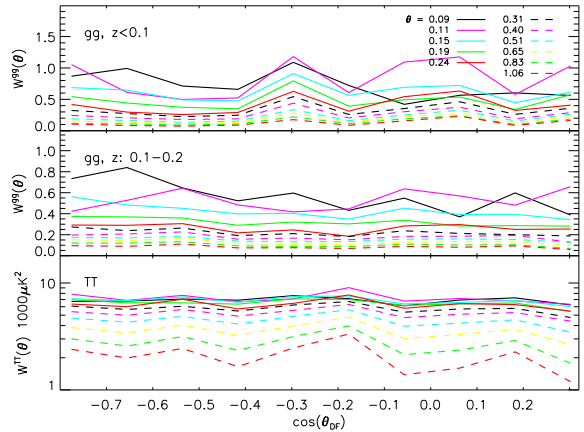}}
\caption{The $\cos(\theta_{\rm DF})$ dependence of galaxy and CMB 
auto-correlations. Top: galaxy auto-correlation in redshift range 
$z<0.1$. Middle: galaxy auto-correlation in redshift reange $0.1<z<0.2$. 
Bottom: CMB auto-correlation.}
\label{autosub}
\end{figure}

One measure of the residual CMB/galaxy contaminations is the isotropy 
of CMB and galaxy distribution. We measure the sky-dependence of the 
galaxy and the CMB temperature auto-correlation in the same subregions. 
The results are shown in figure~\ref{autosub}. None of them show 
statistically significant sky dependence. This implies that the residual
contaminations are well under control and it is unlikely to be the 
cause of the measured anisotropic cross correlation. 

\subsubsection{CMB foregrounds}
\begin{figure} 
\resizebox{\hsize}{!}{\includegraphics{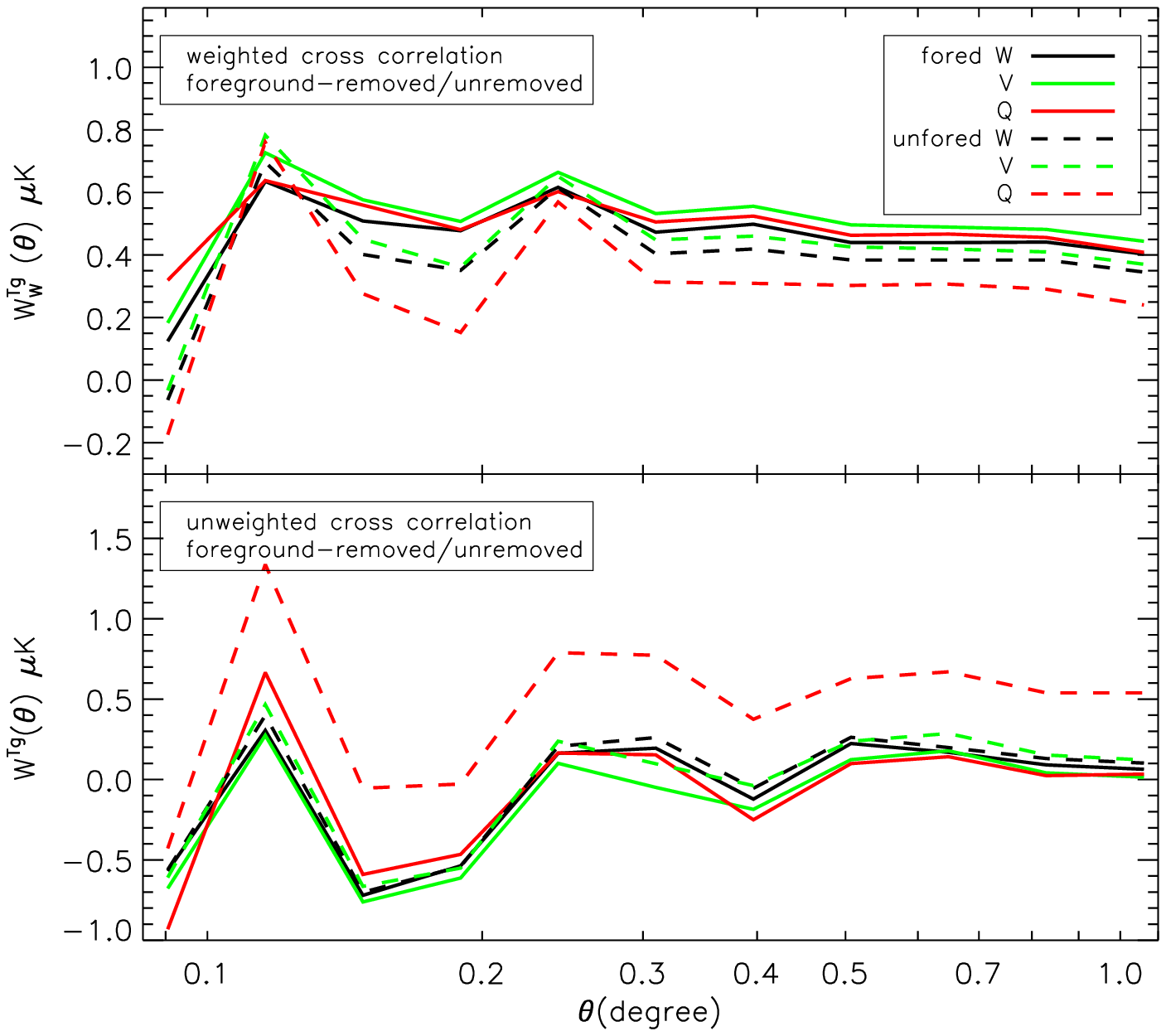}}
\caption{The effect of foreground on weighted (upper) and unweight 
(lower) cross correlation measurement. The black, green and red lines 
show results W, V and Q band respectively.}
\label{foredremove}
\end{figure}

We compared the cross correlation results using the foreground removed 
CMB maps with the ones which use maps without foreground removal, 
the results are shown in figure~\ref{foredremove}. 
As can be seen from the figure, the high frequency band W and V do not 
suffer much from the foreground, while the Q band is contaminated 
badly. After the foreground removal procedure, the frequency dependence 
is greatly suppressed for both the weighted and unweighted cross 
correlations, showing that the foreground contamination has been 
largely cleaned, and its contribution to the cross correlations should 
be small.

\subsection{The dark flow induced kSZ effect?}
We find that the observed cross correlation can not be explained by the
reported dark flow scenario either, mainly due to two reasons. First, 
the observed directional dependence does not match the prediction of
dark flow. As explained earlier, the dark flow induced kSZ effect predicts a unique 
relation $w^{Tg}(\theta|\theta_{\rm DF})\propto \cos\theta_{\rm DF}$. 
However, figure~\ref{gtsub} shows that the measured directional dependence 
is rather complicated and can not be well described by the above
relation. Furthermore, a puzzling issue is that over the range
$\cos(\theta_{\rm DF})>-0.3$, the cross correlations for both redshift 
bins do follow the $\cos(\theta_{\rm DF})$ dipole pattern. This is also 
the case for the $0.1<z<0.2$ bin with $\cos(\theta_{\rm DF})<-0.4$. 
Given the noisiness of the measurements, we do not attempt to quantify 
the statistical significance of these behaviors. Nevertheless, the 
complicated directional dependence suggests that it unlikely, for the 
cross correlations we found, comes from a single dark flow.

Second, to match the measurements, the required dark flow amplitude 
must be much larger than the $\sim 1000$ km/s value reported by
\cite{2010ApJ...712L..81K}.  For quick comparison, in 
figure~\ref{ksz} we plot the theoretical
predictions of kSZ signal coming from dark flow, using method
described in section \ref{theory}.   We adopt a value of $9530\kms$ over
redshift range $z<0.1$ (black dashed line) and  
$6140\kms$ over redshift range $0.1<z<0.2$ (red dashed line). A dark flow of
such large amplitude is unrealistic, since it would cause the kSZ signal of
clusters to be comparable  or even dominate over
the tSZ signal of $\sim 10$ keV clusters and contradict existing SZ observation of clusters
at low redshifts, especially the measurements at the 217 GHz frequency band
(e.g. the Planck intermediate results, \citealt{PlanckSZ2012}). 

Hence we rule out the possibility of  a dark flow as the major source of the
observed anomalous anisotropic WMAP-SDSS cross correlation. However, the
above results do not  rule out a dark flow of much smaller amplitude, such as the
$\sim 1000$ km/s dark flow reported by
\cite{2010ApJ...712L..81K}. Such dark flow would contribute a sub-dominant
fraction to the measured cross correlation and hence do not necessarily
contradict with the measurements. The low signal-to-noise ratio in our
measurement   prohibits further investigation. The upcoming Planck 
project, with better angular resolution and lower instrumental noise, 
could provide us more information on the true origin of the anomalous 
anisotropic correlations.

\section{Summary}   \label{conclusion}

In this paper we report detection of anomalous anisotropic CMB-galaxy cross 
correlations using KIAS-VAGC galaxy catalog of SDSS-DR7 and WMAP7 CMB
data and discuss their implications in the dark flow scenario. 
We measure this anisotropy using a specially weighted cross
correlation, which vanishes for isotropic components. It is detected 
at 3.5$\sigma$ for  redshift bin $z<0.1$ and 3$\sigma$ for
$0.1<z<0.2$. We measured the sky dependence of the cross correlation. 
It  shows  complicated $\cos(\theta_{\rm DF})$ dependence, inconsistent
with the dark flow scenario.
Also, we find that the measured cross correlation is too large to be
caused by the reported $\sim 1000$ km/s dark flow. 
Improvements in CMB measurement by Planck and
improvements in galaxy surveys will 
help us to significantly improve the cross correlation measurements and better
understand the  origin of the measured anomalous cross correlation.

\section*{Acknowledgments}

We thank Yun-Young Choi and Changbom Park for kindly providing the KIAS-VAGC 
galaxy catalog. We acknowledge the use of the LAMBDA archive (http://lambda.
gsfc.nasa.gov). ZL and XC are supported by the National Science Foundation of 
China (NSFC) under grant No.11073024; XC is supported by the John Templeton Foundation; 
PJZ is supported by NSFC under grant
No. 10821302, 10973027, 11025316,  the CAS/SAFEA 
International Partnership Program for  Creative Research Teams (KJCX2-YW-T23)
and National Basic Research Program of China (973 Program) under grant
No.2009CB24901.

Funding for the SDSS and SDSS-II has been provided by the Alfred P. Sloan
Foundation, the Participating Institutions, the National Science
Foundation, the U.S. Department of Energy, the National Aeronautics and
Space Administration, the Japanese Monbukagakusho, the Max Planck 
Society, and the Higher Education Funding Council for England.  
The SDSS Web Site is http://www.sdss.org/.

The SDSS is managed by the Astrophysical Research Consortium for the
Participating Institutions. The Participating Institutions are the
American Museum of Natural History, Astrophysical Institute Potsdam,
University of Basel, Cambridge University, Case Western Reserve University,
University of Chicago, Drexel University, Fermilab, the Institute for
Advanced Study, the Japan Participation Group, Johns Hopkins University,
the Joint Institute for Nuclear Astrophysics, the Kavli Institute for 
Particle Astrophysics and Cosmology, the Korean Scientist Group, the
Chinese Academy of Sciences (LAMOST), Los Alamos National Laboratory,
the Max-Planck-Institute for Astronomy (MPIA), the Max-Planck-Institute
for Astrophysics (MPA), New Mexico State University, Ohio State University,
University of Pittsburgh, University of Portsmouth, Princeton University,
the United States Naval Observatory, and the University of Washington.

\bibliographystyle{hapj}

\bibliography{df}

\end{document}